\documentclass[conference, 10pt]{IEEEtran}
\IEEEoverridecommandlockouts
\usepackage{cite}
\usepackage{amsmath,amssymb,amsfonts}
\usepackage{algorithmic}
\usepackage{algorithm}
\usepackage{graphicx}
\usepackage{textcomp}
\usepackage{xcolor}
\def\BibTeX{{\rm B\kern-.05em{\sc i\kern-.025em b}\kern-.08em
		T\kern-.1667em\lower.7ex\hbox{E}\kern-.125emX}}
\usepackage{booktabs}
\usepackage{multirow}
\usepackage{xcolor}
\usepackage[super]{nth}
\usepackage{tikz}
\newcommand*\circled[1]{\tikz[baseline=(char.base)]{
		\node[shape=circle,draw,inner sep=0.2pt] (char) {#1};}}

\usepackage{soul}
\usepackage{enumitem}
\newcommand{\add}[1]{\textcolor{black}{#1}}

\usepackage{fancyhdr}
\fancypagestyle{firstpage}{
  \fancyhf{}
  \chead{\small To appear at the 40th IEEE/ACM International Conference on Computer-Aided Design (ICCAD), November 2021, Virtual Event.}
  \cfoot{\thepage}
}

\usepackage[none]{hyphenat}

\begin{document}

\title{ReSpawn: Energy-Efficient Fault-Tolerance for \\ Spiking Neural Networks considering \\ Unreliable Memories
\vspace{-0.2cm}
}
	
\author{\IEEEauthorblockN{Rachmad Vidya Wicaksana Putra\IEEEauthorrefmark{1}, Muhammad Abdullah Hanif\IEEEauthorrefmark{2}, Muhammad Shafique\IEEEauthorrefmark{3}}
\IEEEauthorblockA{\IEEEauthorrefmark{1}\IEEEauthorrefmark{2}\textit{Technische Universit\"at Wien (TU Wien)}, Vienna, Austria \\
\textit{\IEEEauthorrefmark{2}\IEEEauthorrefmark{3}New York University Abu Dhabi (NYUAD)}, Abu Dhabi, United Arab Emirates \\
Email: \{rachmad.putra, muhammad.hanif\}@tuwien.ac.at,
muhammad.shafique@nyu.edu}
\vspace{-0.9cm}
}

\maketitle
\thispagestyle{firstpage}

\begin{abstract}
Spiking neural networks (SNNs) have shown a potential for having low energy with unsupervised learning capabilities due to their biologically-inspired computation. 
However, they may suffer from accuracy degradation if their processing is performed under the presence of hardware-induced faults in memories, which can come from manufacturing defects or voltage-induced approximation errors. 
Since recent works still focus on the fault-modeling and random fault injection in SNNs, the impact of memory faults in SNN hardware architectures on accuracy and the respective fault-mitigation techniques are not thoroughly explored. 
Toward this, we propose ReSpawn, a novel framework for mitigating the negative impacts of faults in both the off-chip and on-chip memories for resilient and energy-efficient SNNs. 
The key mechanisms of ReSpawn are: (1) analyzing the fault tolerance of SNNs; and (2) improving the SNN fault tolerance through (a) fault-aware mapping (FAM) in memories, and (b) fault-aware training-and-mapping (FATM). 
If the training dataset is not fully available, FAM is employed through efficient bit-shuffling techniques that place the significant bits on the non-faulty memory cells and the insignificant bits on the faulty ones, while minimizing the memory access energy.
Meanwhile, if the training dataset is fully available, FATM is employed by considering the faulty memory cells in the data mapping and training processes.
The experimental results show that, compared to the baseline SNN without fault-mitigation techniques, ReSpawn with a fault-aware mapping scheme improves the accuracy by up to 70\% for a network with 900 neurons without retraining.
\end{abstract}


\vspace{-0.1cm}
\section{Introduction}
\label{Sec_intro}

\begin{figure}[t]
\centering
\includegraphics[width=\linewidth]{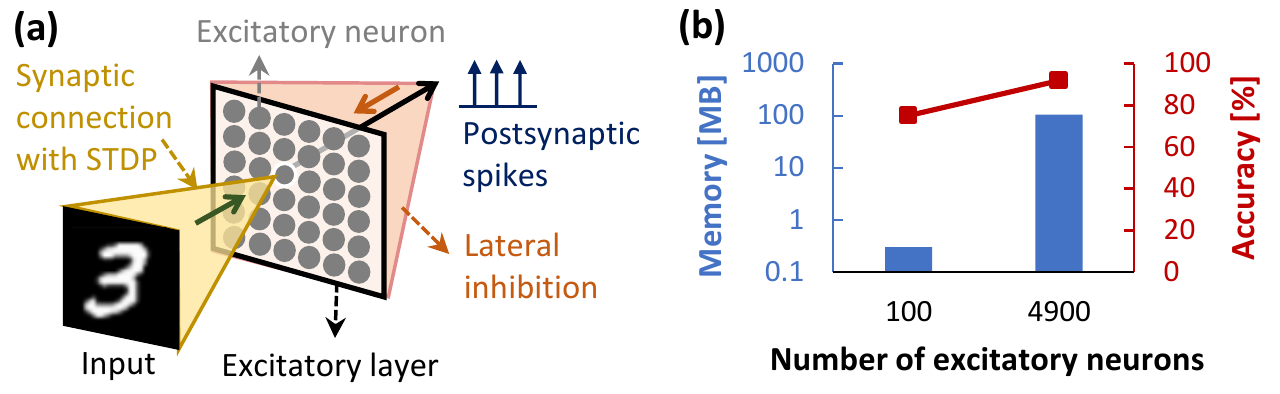}
\vspace{-0.7cm}
\caption{(a) SNN architecture considered in this work, i.e., a single-layer fully- connected network. (b) An SNN model with a larger model size typically achieves higher accuracy than the smaller ones.}
\label{Fig_ObserveTopoMemAcc}
\vspace{-0.6cm}
\end{figure}

Spiking neural networks (SNNs) have shown promising performance by achieving high accuracy with low energy consumption in an unsupervised learning scenario \cite{Ref_Pfeiffer_DLSNN_FNINS18,Ref_Tavanaei_DLSNN_Neunet18,Ref_Putra_FSpiNN_TCAD20,Ref_Putra_SpikeDyn_arXiv21}.
Currently, a large-sized SNN model is more favorable than the smaller ones since it can achieve higher accuracy, but at the cost of higher memory footprint and energy consumption \cite{Ref_Putra_FSpiNN_TCAD20}, as illustrated in Fig.~\ref{Fig_ObserveTopoMemAcc}. 
To address these challenges, SNN accelerators have been developed to improve the performance and energy-efficiency of SNN-based applications \cite{Ref_Akopyan_TrueNorth_TCAD15,Ref_Roy_PEASE_ISLPED17,Ref_Sen_ApproxSNN_DATE17,Ref_Davies_Loihi_MM18,Ref_Frenkel_ODIN_TBCAS19,Ref_Frenkel_MorphIC_TBCAS19}. 
However, these SNN accelerators may suffer from accuracy degradation when SNN processing is performed under the presence of \textit{hardware-induced faults in memories}, which can come from the following sources.
\begin{itemize}[leftmargin=*]
    \item \textit{Manufacturing defects}: The imperfections in the chip fabrication process can cause defects in memory cells, which degrades the cell functionality in the form of faults, and thereby reducing the yield of chips \cite{Ref_Koren_Yield_IEEE98}. 
    \item \textit{Voltage-induced approximation errors}: The operational voltage of memories can be reduced to decrease the power and energy, at the cost of increased fault rates \cite{Ref_Chang_Voltron_POMACS17}\cite{Ref_Ganapathy_UnreliableMem_DAC15}.
\end{itemize}   

\textbf{Targeted Research Problem:} \textit{If and how can we mitigate the negative impacts of faults in the off-chip and on-chip weight memories on the accuracy, thereby improving the fault tolerance of SNNs and maintaining high accuracy}. 

The efficient solution to this problem will enable reliable SNN processing in the presence of unreliable memories for energy-constrained devices, such as IoT-Edge. 
Furthermore, this will also improve the yield and reduce the per-unit-cost of the SNN accelerator.

\vspace{-0.1cm}
\subsection{State-of-the-Art and Their Limitations}
\label{Sec_SOTA}
\vspace{-0.1cm}

Standard fault-tolerance techniques for VLSI circuits (such as DMR \cite{Ref_Vadlamani_DMR_DATE20}, TMR \cite{Ref_Lyons_TMR_IBM62}, and ECCs \cite{Ref_Sze_ECCs_USPatent00})\footnote{DMR: Dual Modular Redundancy, TMR: Triple Modular Redundancy, and ECCs: Error Correcting Codes.} are not effective for improving the resiliency of SNN systems, as they need redundant hardware and/or execution which incurs high overhead. 
To address this, state-of-the-art works have proposed fault tolerance for SNNs, which can be categorized as follows.
\begin{itemize}[leftmargin=*]
    \item \textbf{Fault modeling of SNNs:} It discusses (1) a set of possible faults that can affect SNN components, such as neurons and synapses \cite{Ref_Vatajelu_ReliabilitySNN_VTS19}, and (2) the fault modeling for analog neuron hardware \cite{Ref_Sayed_NeuronFaultModel_IOLTS20} as well as its fault-tolerance strategy \cite{Ref_Spyrou_NeuronFT_DATE21}.  
    \item \textbf{Studying the impacts of faults on accuracy:} It explores and discusses the impacts of bit-flips in weights \cite{Ref_Venceslai_NeuroAttack} and full synapse failure (i.e., synapse removal) \cite{Ref_Schuman_ResilinceSNN_IJCNN20}\add{\cite{Ref_Rastogi_AstrocytesSTDP_FNINS21}} on the accuracy, considering different fault rates with random distribution. 
    \item \add{\textbf{Retraining-based fault mitigation:} Work \cite{Ref_Rastogi_AstrocytesSTDP_FNINS21} incorporates additional components (i.e., astrocyte units) to the network for enhancing the retraining process for mitigating the faults.}  
\end{itemize}

\textbf{Limitations:} 
These state-of-the-art still focus on SNN fault modeling and fault injection without considering the SNN hardware architectures. 
Therefore, \textit{the impact of bit-level faults in the off-chip and the on-chip memories on the accuracy, and the respective fault-mitigation techniques, are still unexplored}.
\add{Moreover, \textit{current fault mitigation techniques still rely on the costly additional components and retraining process}}.

To understand the characteristics of the SNN fault tolerance, we present an experimental case study in the following section.

\subsection{Motivational Case Study and Research Challenges}
\label{Sec_MotivationChallenges}

\begin{figure}[t]
\centering
\includegraphics[width=0.95\linewidth]{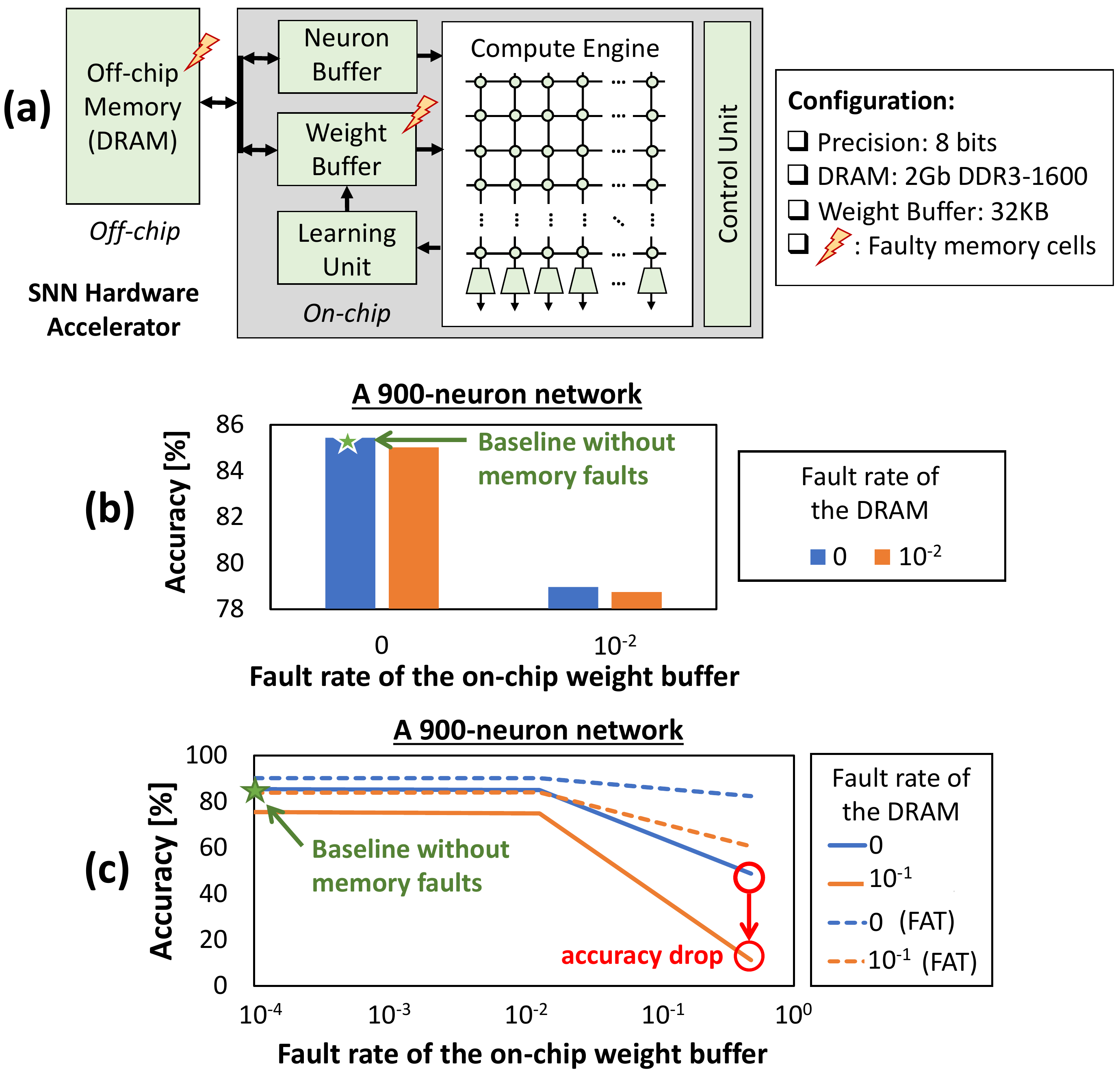}
\vspace{-0.3cm}
\caption{(a) The SNN accelerator considered in this work with faults in the DRAM and the weight buffer. We employ 8-bit precision, a 2Gb DDR3-1600 DRAM, a 32KB weight buffer, and uniform random distribution of faults on each bank of the DRAM and the weight buffer in the form of bit-flip. (b) Significance of faults in the DRAM and the weight buffer on the accuracy. (c) Increasing fault rates in weight memories lead to accuracy degradation, and fault-aware training (FAT) can improve the SNN fault tolerance.}
\label{Fig_ObserveArchAccDrop}
\vspace{-0.4cm}
\end{figure}

To understand the impact of faults (from the manufacturing defects and the reduced-voltage operations in memories) on the accuracy, we perform experiments that explore different fault rates in the DRAM-based off-chip memory and the SRAM-based on-chip weight memory (weight buffer), while considering the typical architecture of SNN accelerators, as shown in Fig.~\ref{Fig_ObserveArchAccDrop}(a).  
Further details on the experimental setup are presented in Section~\ref{Sec_EvalMethod}.
 
From the experimental results in Fig.~\ref{Fig_ObserveArchAccDrop}(b)-(c), we make the following key observations. 
\begin{itemize}[leftmargin=*]
    \item Different fault rates in the DRAM and the weight buffer cause an SNN system to obtain different accuracy scores.
    \item Faults in the weight buffer have a relatively higher impact on the accuracy degradation than the DRAM, since its size is significantly smaller than DRAM, and thereby having a higher probability to affect more weights. 
    \item Fault-aware training (FAT) with progressive fault injection for neural networks \cite{Ref_Koppula_EDEN_MICRO19}\cite{Ref_Putra_SparkXD_arXiv21} can improve the SNN fault tolerance while incurring high training time and energy, as it considers a wide range of fault rates for the injection.
\end{itemize}

\textbf{Research Challenges:} 
The above observations expose key challenges that need to be solved for addressing the targeted research problem, as discussed in the following. 
\begin{itemize}[leftmargin=*]
    \item \textit{The fault-mitigation technique should minimize the impacts of faults in both, the DRAM and the weight buffer}, thereby improving the SNN fault tolerance.
    \item \textit{It should employ a technique that does not rely on retraining}, as retraining needs a full training dataset, which may not be available due to restriction policies (e.g., a company releases an SNN model, but makes the training dataset unavailable).
    \item \textit{It should incur low energy overhead at run-time}, compared to the baseline (without fault-mitigation technique) to enable energy-efficient SNN applications.
\end{itemize}

\renewcommand{\headrulewidth}{0pt}
\vspace{-0.1cm}
\subsection{Our Novel Contributions}
\label{Sec_Novelty}
\vspace{-0.1cm}

\begin{figure}[hbtp]
\vspace{-0.3cm}
\centering
\includegraphics[width=0.9\linewidth]{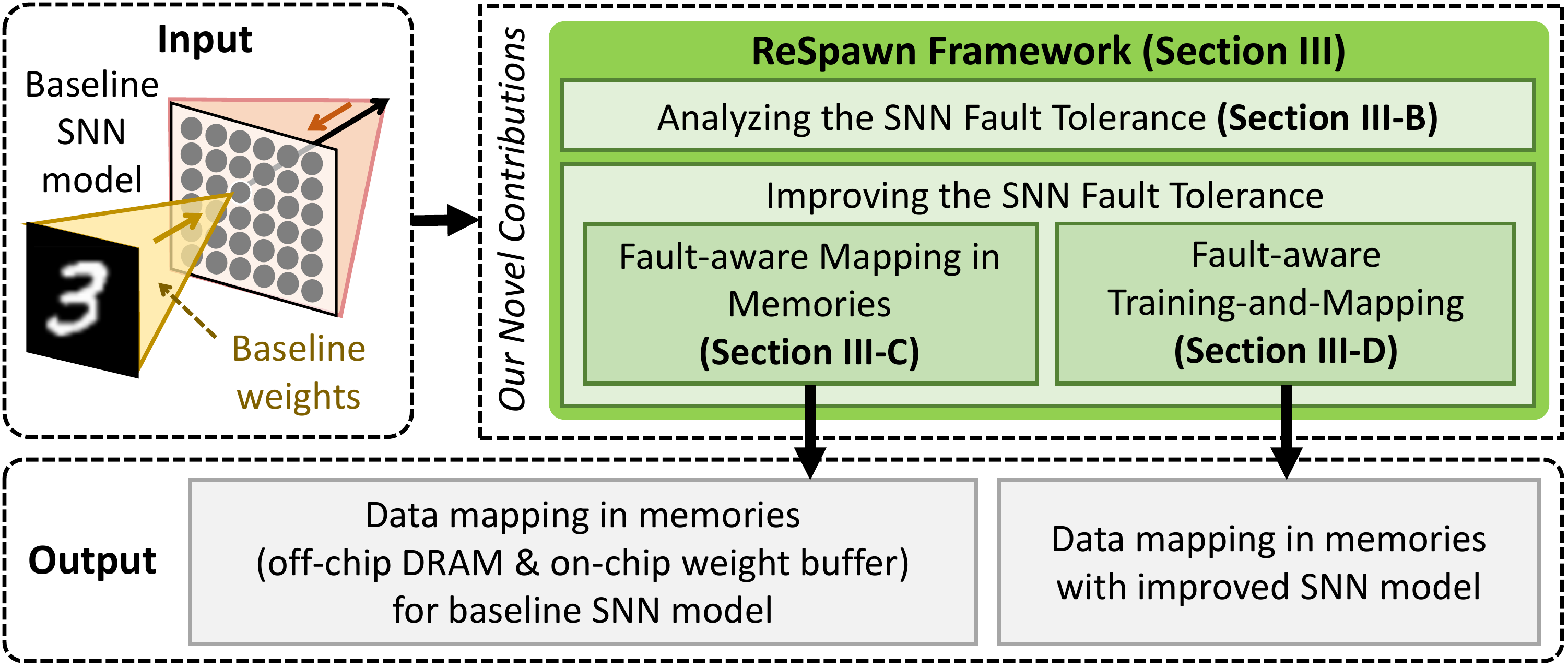}
\vspace{-0.3cm}
\caption{An overview of our novel contributions (shown in the green boxes).}
\label{Fig_ReSpawn_Novelty}
\vspace{-0.4cm}
\end{figure}

To address the above challenges, we propose \textbf{ReSpawn framework} that enables \textit{energy-efficient fault-tole\underline{R}anc\underline{e} for \underline{Sp}iking neur\underline{a}l net\underline{w}orks considering u\underline{n}reliable memories}.
\add{To the best of our knowledge, this work is the first effort that mitigates the negative impacts of faults in the off-chip and on-chip weight memories of SNN accelerators.}
Following are the key steps of the ReSpawn framework (see Fig.~\ref{Fig_ReSpawn_Novelty}): 
\begin{enumerate}[leftmargin=*] 
  \item \textbf{Analyzing the Fault Tolerance of the SNN Model} to characterize the accuracy values under the given fault rates. 
  It is performed by adjusting the fault rates in the memories, while checking the obtained accuracy.
  \item \textbf{Improving the Fault Tolerance of the SNN model} whose strategies depend on the availability of the training dataset.
  \begin{itemize}
      \item \textit{If the training dataset is not fully available,} \textbf{the Fault- aware Mapping (FAM)} is employed through simple bit- shuffling techniques, that prioritize placing the bits with higher significance on the non-faulty memory cells. 
      \item \textit{If the training dataset is fully available,} \textbf{the Fault-aware Training-and-Mapping (FATM)} is employed by including the information of the faulty memory cells in the data mapping and training processes. Here, the data mapping strategy follows the proposed FAM technique.
  \end{itemize}
\end{enumerate}

\textbf{Key Results:} 
We evaluated our ReSpawn framework for accuracy and energy, using Python-based simulations on the GPGPU. 
The experimental results show that ReSpawn with fault-aware mapping improves the accuracy by up to 70\% for a 900-neuron network without retraining on the MNIST dataset. 

\section{Preliminaries}
\label{Sec_Prelim}

\subsection{Spiking Neural Networks (SNNs)}
\label{Sec_Prelim_SNNs}

The components of an SNN model consist of spike coding, neuron model, network architecture, and learning rule. 
Spike coding represents the input into spikes.  
Different types of spike coding have been proposed, such as rate, rank-order, phase, burst, and time-to-first spike \cite{Ref_Gautrais_SpikeCoding_Bio98, Ref_Thorpe_RankOrder_Springer98,Ref_Park_BurstSNN_DAC19,Ref_Park_T2FSNN_DAC20}, and here we select the rate coding due to its robustness for data representation. 
For the neuron model, we choose the Leaky Integrate-and-Fire (LIF) as it has low complexity \cite{Ref_Putra_QSpiNN_arXiv21}. 
Its membrane potential increases if a presynaptic spike comes, and it decreases if no presynaptic spike is observed. 
If the potential reaches the threshold ($V_{th}$), a postsynaptic spike is generated, and afterward, it is back to the reset potential ($V_{reset}$).
We consider the network architecture of Fig.~\ref{Fig_ObserveTopoMemAcc}(a), i.e., a single-layer fully-connected network, since it shows the state-of-the-art accuracy for the unsupervised learning scenario \cite{Ref_Putra_FSpiNN_TCAD20}. 
In such an architecture, the spikes generated from each neuron are passed to other neurons for providing inhibition, which promotes competition among neurons to recognize the given input. 
We employ the spike-timing-dependent plasticity (STDP) for the learning rule, due to its capability for learning input features under unsupervised settings.
\add{We consider the unsupervised SNNs since they can learn features from the unlabeled data, which is highly desired for real-world applications (gathering unlabeled data is easier and cheaper than labeled ones) \cite{Ref_Putra_FSpiNN_TCAD20}}.

\subsection{Fault Modeling for Memories}
\label{Sec_ErrorModelMemories}

We focus on the fault modeling for the DRAM and the SRAM weight buffer, since we aim at accurately exploring the impacts of hardware-induced faults in weight memories across the hierarchy of an SNN accelerator, as shown in Fig.~\ref{Fig_ObserveArchAccDrop}(a). 
These faults can come from \textit{manufacturing defects}, due to the imperfections in the fabrication process \cite{Ref_Koren_Yield_IEEE98}\cite{Ref_Cesar_ReviewFTinNN_Access17,Ref_Hanif_RobustML_IOLTS18,Ref_Zhang_RobustML_DAC19,Ref_Shafique_RobustML_DnT20}, and \textit{reduced-voltage operation}, which is performed for decreasing the operational power/energy \cite{Ref_Chang_Voltron_POMACS17}\cite{Ref_Ganapathy_UnreliableMem_DAC15}. 

\textbf{Faults from Manufacturing Defects:} 
The SNN hardware accelerators are fabricated using a sophisticated manufacturing process. 
Hence, there is a chance of imperfections that result in defects in the fabricated chips.  
Moreover, the technology scaling, which is employed for improving the performance and efficiency of the chips, may increase fault rates related to both, permanent faults (e.g., stuck-at faults) and transient faults (e.g., bit-flip) at random locations of a chip.
Therefore, the faults from manufacturing defects can be modeled using a uniform random distribution, which has also been considered in previous works \cite{Ref_Zhang_PermanentFaults_VTS18}\cite{Ref_Abdullah_SalvageDNN_RSTA20}.

\begin{figure}[hbtp]
\vspace{-0.2cm}
\centering
\includegraphics[width=\linewidth]{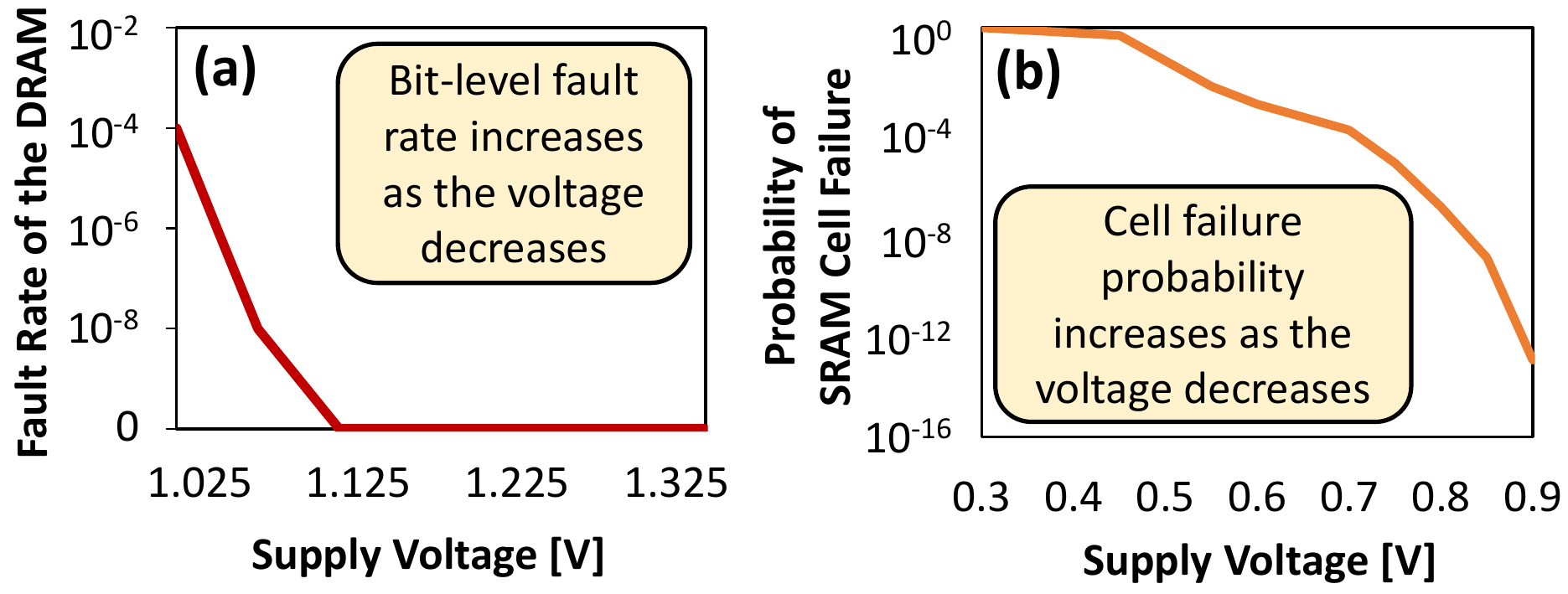}
\vspace{-0.7cm}
\caption{(a) DRAM fault rates and the corresponding
DRAM voltage values, based on the study of \cite{Ref_Chang_Voltron_POMACS17}. (b) SRAM cell failure probability ($P_{cell}$) and the corresponding
SRAM voltage values for a 28nm CMOS technology, based on the study of \cite{Ref_Ganapathy_UnreliableMem_DAC15}. The yield of non-faulty cells is defined as $Y = (1-P_{cell})^M$ with $M$ denotes the total memory bit-cells.}
\label{Fig_DRAMnSRAMfaultrates}
\vspace{-0.4cm}
\end{figure}

\textbf{Faults from Reduced-Voltage Operations:} 
The reduction of operational voltage is a widely-used approach to reduce the operational power/energy of DRAM and SRAM-based buffer, at the cost of increased fault rates, as shown in Fig.~\ref{Fig_DRAMnSRAMfaultrates}.
For DRAM, we follow the fault model from \cite{Ref_Koppula_EDEN_MICRO19}, i.e., the faults are modeled by considering the \textit{weak cells} (i.e., cells that fail when the DRAM voltage is reduced), and the probability of a fault in any weak cell.
Here, the faults have a uniform random distribution across a DRAM bank. 
Meanwhile, for SRAM, we follow the fault model from \cite{Ref_Ganapathy_UnreliableMem_DAC15}, i.e., the faults have a uniform random distribution across an SRAM bank.
The selection of the uniform random distribution as the fault model for DRAM and SRAM, is motivated by the following reasons: 
(1) it produces faults with high similarity to the real reduced-voltage DRAM \cite{Ref_Koppula_EDEN_MICRO19} and the real reduced-voltage SRAM \cite{Ref_Ganapathy_UnreliableMem_DAC15}; 
and (2) it offers fast software-based fault injection.

\section{ReSpawn Framework}
\label{Sec_ReSpawn}

\subsection{Overview}
\label{Sec_ReSpawn_Overview}

\begin{figure}[hbtp]
\centering
\includegraphics[width=\linewidth]{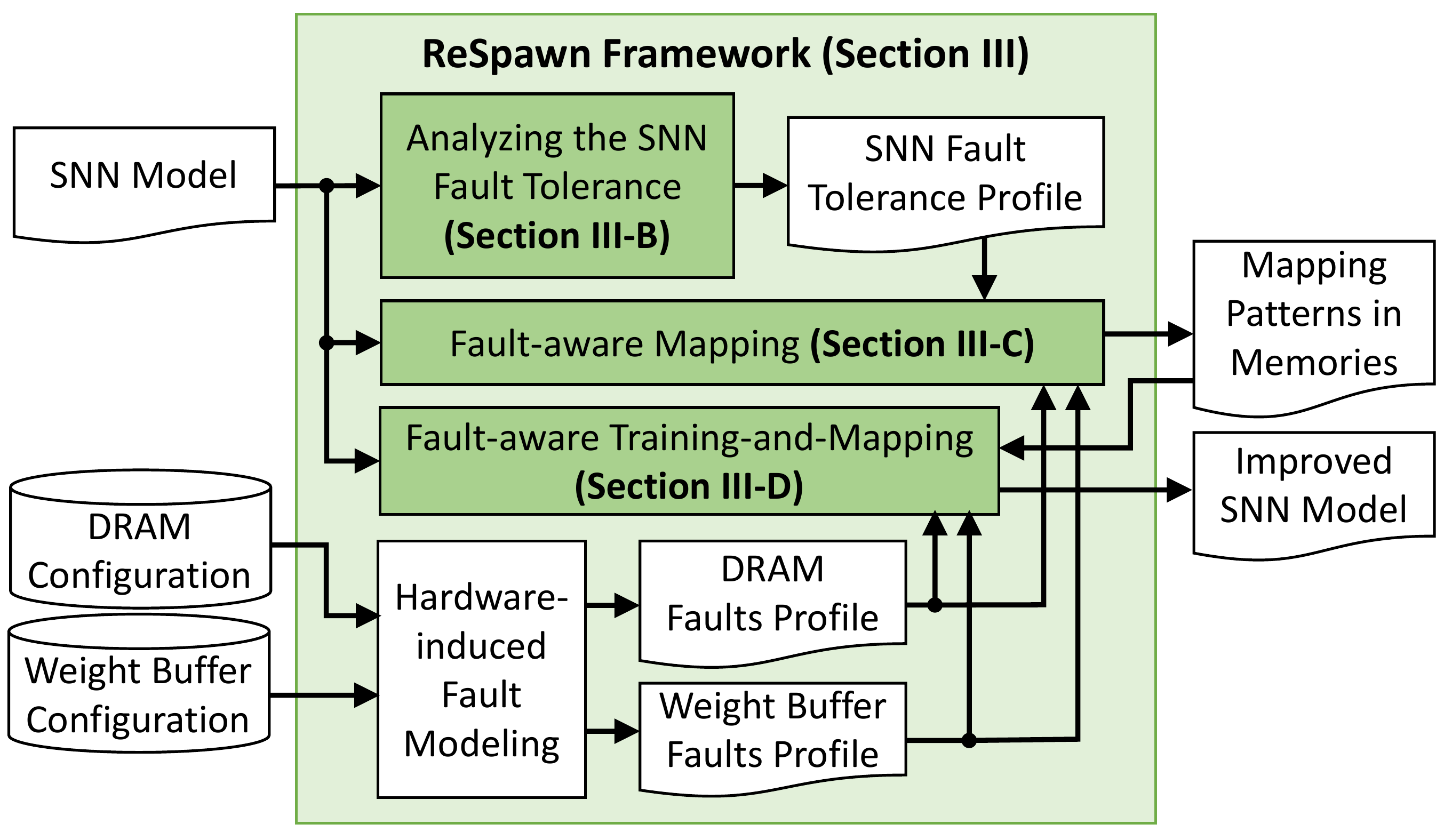}
\vspace{-0.6cm}
\caption{An overview of our ReSpawn framework, whose novel mechanisms are shown in the green boxes.}
\label{Fig_ReSpawn_Overview}
\vspace{-0.3cm}
\end{figure}

We propose the ReSpawn framework for energy-efficient fault-tolerance for SNNs processing on unreliable off-chip and on-chip weight memories. 
The key steps of our ReSpawn are shown in Fig.~\ref{Fig_ReSpawn_Overview}, and discussed in the following sections.
\begin{enumerate}[leftmargin=*] 
  \item \textbf{Analyzing the SNN Fault Tolerance (Section~\ref{Sec_ReSpawn_SNNftAnalysis}):} 
  It aims at understanding the interaction between the fault rates and the accuracy, by exploring different combinations of fault rates in DRAM and weight buffer, while observing the accuracy scores.
  This information is then leveraged for improving the SNN fault tolerance. 
  \item \textbf{Improving the SNN Fault Tolerance} through different strategies, depending on the availability of training dataset.
  \begin{itemize}
      \item \textbf{Fault-aware Mapping (Section~\ref{Sec_ReSpawn_FM}):} 
      \textit{This strategy is performed if the training dataset is not fully available.}
      It employs efficient bit-shuffling techniques, that place the significant bits on the non-faulty memory cells and the insignificant bits on the faulty ones.
      We propose two FAM techniques to offer  accuracy-energy trade-offs.
      \begin{itemize}
          \item \textbf{FAM1:} It considers the fault map from each memory as an individual fault map, and devises a mapping pattern for each fault map.
          \item \textbf{FAM2:} It merges multiple fault maps from different memories to an integrated fault map, and devises a mapping pattern for it accordingly.
      \end{itemize}
      \item \textbf{Fault-aware Training-and-Mapping  (Section \ref{Sec_ReSpawn_FMT}):} 
      \textit{This strategy is performed if the training dataset is fully available.}
      It uses the information of the faulty memory cells in the data mapping and training processes. 
      Here, the data mapping strategy also follows the proposed FAM techniques (i.e., FAM1 and FAM2).
  \end{itemize}
\end{enumerate}

\subsection{SNN Fault Tolerance Analysis}
\label{Sec_ReSpawn_SNNftAnalysis}

\begin{figure}[hbtp]
\vspace{-0.3cm}
\centering
\includegraphics[width=\linewidth]{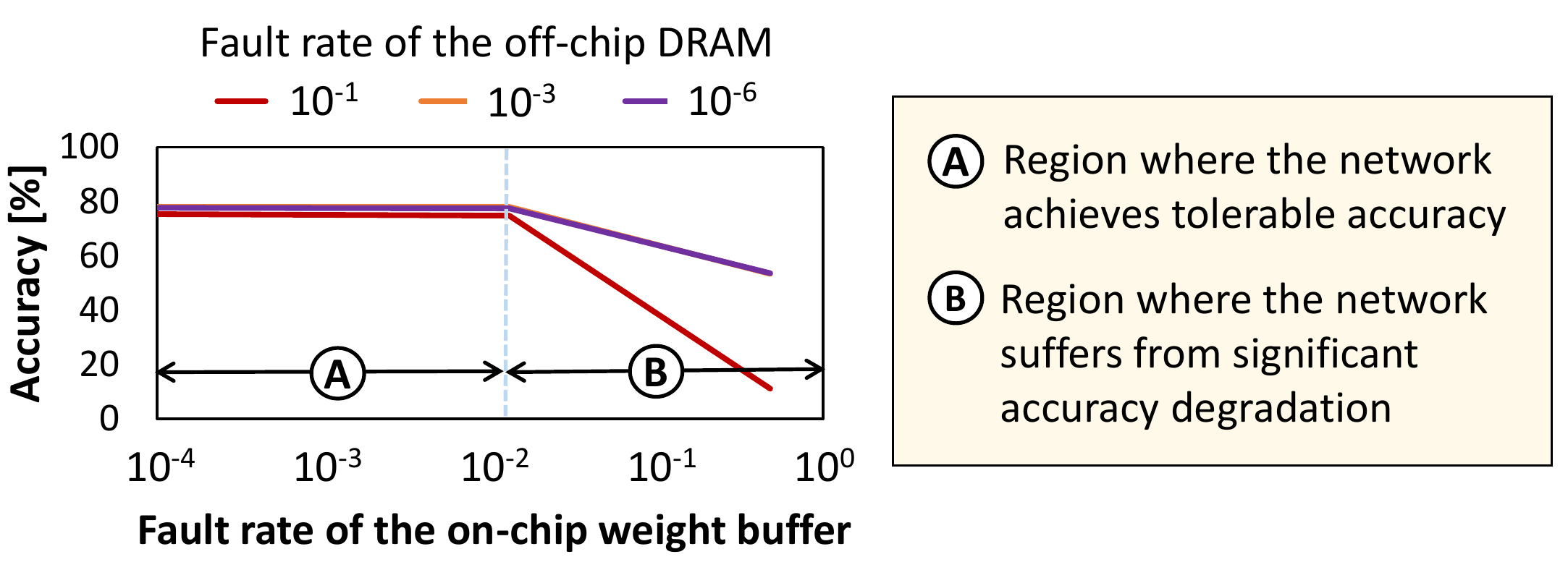}
\vspace{-0.7cm}
\caption{The experimental results for a 900-neuron network, considering different fault rates for DRAM and weight buffer.}
\label{Fig_FaultToleranceAnalysis}
\vspace{-0.2cm}
\end{figure}

Understanding the fault tolerance of the given SNN model is important, because the information from the analysis will be beneficial, especially for performing efficient fault-mitigation techniques. 
Therefore, \textit{our ReSpawn framework analyzes the fault tolerance of the SNN model to observe the interaction between memory faults and accuracy}. 
It is performed by exploring different combinations of fault rates in the DRAM and the weight buffer, while observing the obtained accuracy. 
For instance, if we consider a network with 900 neurons, our ReSpawn will explore different combinations of fault rates in DRAM and weight buffer, and the experimental results are shown in Fig.~\ref{Fig_FaultToleranceAnalysis}. 
These results show two different regions, i.e., where fault rates in memories cause the network to achieve acceptable accuracy, as shown by label-\circled{A}, and where fault rates in memories cause the network to suffer from notable accuracy degradation, as shown by label-\circled{B}. 
These regions provide insights regarding the tolerable fault rates that should be considered to effectively improve the SNN fault tolerance. 

\subsection{Fault-aware Mapping (FAM)}
\label{Sec_ReSpawn_FM}

Faulty cells in memories that come from manufacturing defects and reduced-voltage operations, can be characterized at design time. 
Therefore, their locations are known before the deployment.
\textit{ReSpawn framework leverages the information of faulty cells in DRAM and weight buffer to effectively map the weights on memory fabrics, thereby minimizing the impact of faulty cells on the significant bits}. 
It is performed through fault-aware mapping (FAM), that employs simple bit-shuffling techniques for placing the significant bits on the non-faulty memory cells and the insignificant bits on the faulty ones. 

\begin{figure}[hbtp]
\vspace{-0.2cm}
\centering
\includegraphics[width=\linewidth]{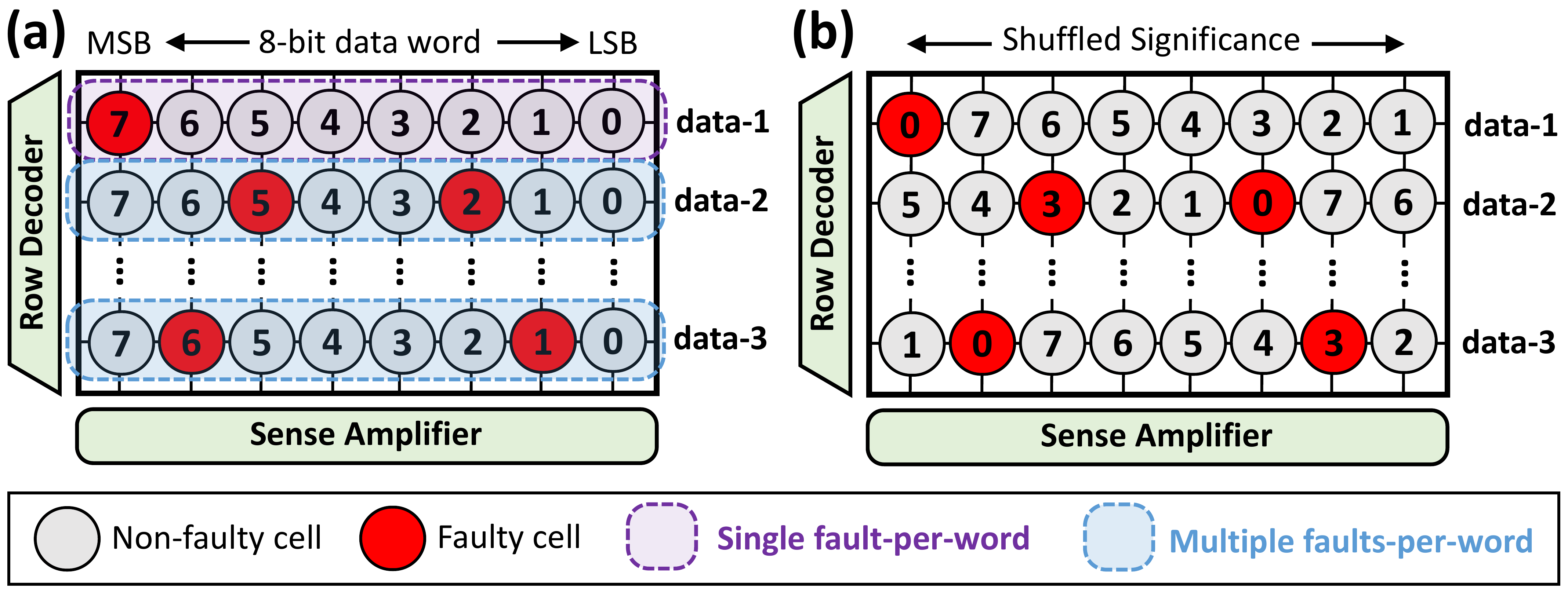}
\vspace{-0.7cm}
\caption{(a) Illustration of possible locations of faulty cells in memories. (b) The proposed bit-shuffling technique, which is based on the right circular-shift.}
\label{Fig_FaultyCells_FM}
\vspace{-0.2cm}
\end{figure}

Furthermore, we observe that, a data word may have a single faulty bit or multiple faulty bits, depending on whether this word occupies a memory segment that has a single faulty cell or multiple faulty cells, as illustrated in Fig.~\ref{Fig_FaultyCells_FM}(a).
Therefore, we propose a mapping strategy that can address both, the single fault-per-word and multiple faults-per-word scenarios.

The proposed memory mapping strategy is performed by the following steps, which are also illustrated in Fig.~\ref{Fig_FaultyCells_FM}(b).
\begin{itemize}[leftmargin=*]
    \item \textbf{Step-1:} \textit{Identifying the faulty cells in the given memories.}
    This step aims at obtaining the information of faulty cells in each memory, such as fault rate and fault map.
    \add{The faulty cells in the on-chip buffer from manufacturing defects can be detected using the standard post-fabrication testing~\cite{Ref_Zhang_PermanentFaults_VTS18}, and the faulty cells in the DRAM can be detected through measurements, e.g., using SoftMC tool~\cite{Ref_Hassan_SoftMC_HPCA17}.
    Meanwhile, the faulty cells from reduced-voltage operations can also be detected through measurements on the DRAM \cite{Ref_Hassan_SoftMC_HPCA17} and on the on-chip buffer \cite{Ref_Ganapathy_UnreliableMem_DAC15}.
    In this manner, collecting the faulty cell information is feasible as it follows the standard post-fabrication testing and measurements}.
    \item \textbf{Step-2:} \textit{Identifying the maximum fault rate allowed in a data word.} 
    This step aims at determining which memory cells, that can be used for storing a data word, by considering the interaction between fault rates and accuracy from SNN fault tolerance analysis in Section~\ref{Sec_ReSpawn_SNNftAnalysis}.
    For instance, we allow a maximum of 2 faulty bits for an 8-bit data word.  
    \item \textbf{Step-3:} \textit{Identifying the memory segment with the highest number of subsequent non-faulty cells for storing a data word.}
    It aims at maximizing the possibility of placing the significant bits on the non-faulty cells. 
    Therefore, we also examine the corner case (i.e., the left-most memory cell with the right-most memory cell) as possible subsequent non-faulty cells, as shown in the second row of Fig.~\ref{Fig_FaultyCells_FM}(b).  
    \item \textbf{Step-4:} \textit{Performing circular-shift technique for each data word.} 
    It aims at efficiently performing bit-shuffling. 
    Here, we always employ right circular-shift to simplify the control.  
\end{itemize}

\vspace{0.1cm}
Since the FAM technique leverages the information of fault maps from multiple memories, we propose two variants of FAM techniques to offer different accuracy-energy trade-offs, which are discussed in the following. 

\vspace{0.1cm}
\subsubsection{FAM for Individual Fault Map (FAM1)}
\label{Sec_ReSpawn_FM1}

\begin{figure}[hbtp]
\vspace{-0.4cm}
\centering
\includegraphics[width=\linewidth]{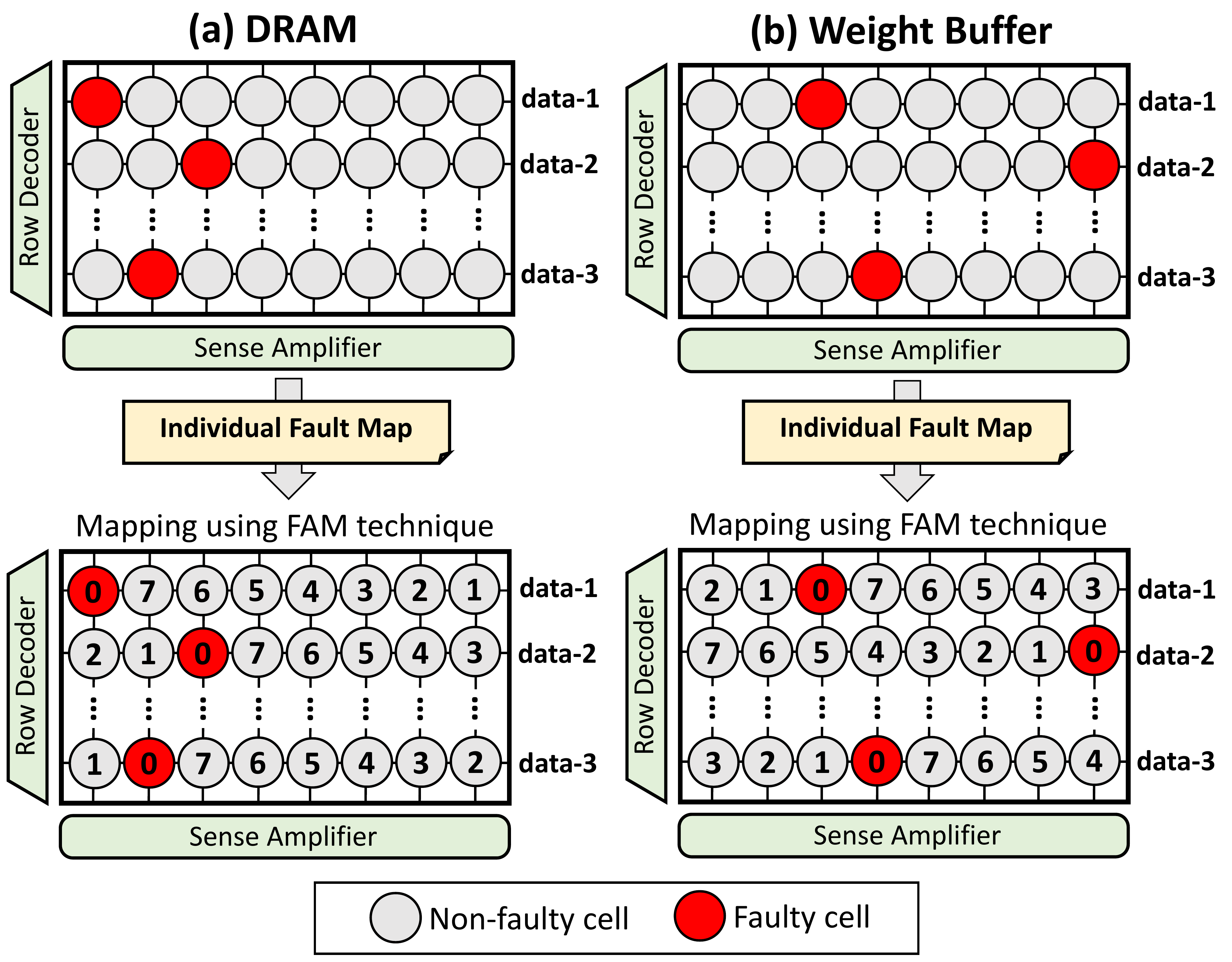}
\vspace{-0.7cm}
\caption{(a) For DRAM, the FAM1 only considers the DRAMs' fault map. (b) For weight buffer, the FAM1 only considers the buffers' fault map.}
\label{Fig_FaultyCells_FM1}
\vspace{-0.3cm}
\end{figure}

This technique considers an individual fault map from each memory (i.e., DRAM or weight buffer). 
Therefore, the FAM1 devises multiple mapping patterns, i.e., one pattern for DRAM, and another one for weight buffer, as illustrated in Fig.~\ref{Fig_FaultyCells_FM1}. 
This FAM1 technique offers high resiliency against faults from each memory, as each mapping pattern minimizes the negative effect of faults on the significant bits.
However, it needs to perform a specialized data mapping for each memory.

\vspace{0.1cm}
\subsubsection{FAM for Integrated Fault Map (FAM2)}
\label{Sec_ReSpawn_FM2}

\begin{figure}[hbtp]
\vspace{-0.3cm}
\centering
\includegraphics[width=\linewidth]{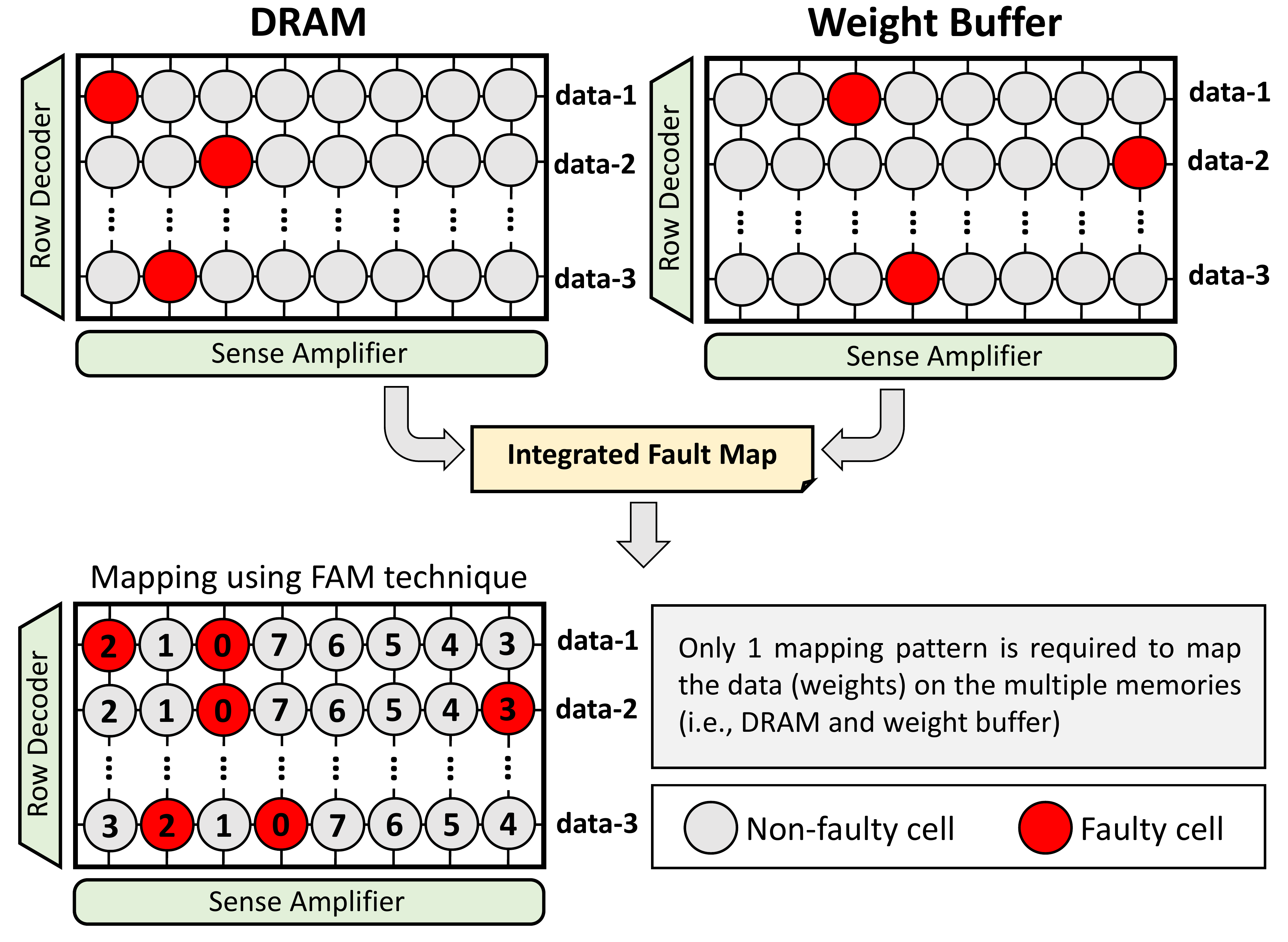}
\vspace{-0.7cm}
\caption{The FAM2 technique considers the integrated fault map for devising the mapping pattern for both, DRAM and weight buffer.}
\label{Fig_FaultyCells_FM2}
\vspace{-0.2cm}
\end{figure}

This technique merges multiple fault maps from multiple memories (i.e., DRAM and weight buffer) as an integrated fault map. 
Hence, the FAM2 only devises a single mapping pattern for both, DRAM and weight buffer, as illustrated in Fig.~\ref{Fig_FaultyCells_FM2}. 
This FAM2 technique potentially offers better efficiency 
than the FAM1, due to its simpler mapping technique.
However, it has less resiliency than the FAM1 as the generated mapping pattern may be sub-optimal for each memory, because some insignificant bits may be placed in non-faulty cells and some significant bits in faulty ones.

ReSpawn also considers optimizing the energy of DRAM and SRAM buffer accesses to maximize the energy efficiency potential, since memory accesses typically dominate the total energy of SNN processing \cite{Ref_Krithivasan_SpikeBundle_ISLPED19}.  
The DRAM mapping is performed by maximizing the DRAM row buffer hits \cite{Ref_Ghose_DRAMworkload_Sigmetrics19}, multi-bank burst feature \cite{Ref_Putra_ROMANet_TVLSI21}, and subarray-level parallelism \cite{Ref_Kim_SALP_ISCA12}\cite{Ref_Putra_DRMap_DAC20}, while considering the proposed FAM techniques (the algorithm is presented in Alg.~\ref{Alg_DRAMmapping}). 
Meanwhile, the SRAM buffer mapping is performed by maximizing the bank-level parallelism \cite{Ref_Putra_ROMANet_TVLSI21} while considering the proposed FAM techniques (the algorithm is presented in Alg.~\ref{Alg_SRAMmapping}). 

\begin{algorithm}[hbtp]
\footnotesize
\caption{\color{black} The proposed mapping for a DRAM chip}
\label{Alg_DRAMmapping}
\begin{algorithmic}[1]
\renewcommand{\algorithmicrequire}{\textbf{INPUT:}}
\renewcommand{\algorithmicensure}{\textbf{OUTPUT:}}
\REQUIRE \textbf{(1)} DRAM ($DRAM$), number of bank-per-chip ($D_{ba}$), number of subarray-per-bank ($D_{su}$), number of row-per-subarray ($D_{ro}$), number of column-per-row ($D_{co}$);\\
\textbf{(2)} Fault rate of a DRAM column ($Drate\_col$), maximum tolerable fault rate for a DRAM column ($Drate\_col_{max}$);\\
\textbf{(3)} Weight bits ($weight\_b$); \\
\textbf{(4)} Fault-aware mapping ($FAM$); // either FAM1 or FAM2\\
\ENSURE DRAM ($DRAM$); \\
\vspace{0.1cm}
\renewcommand{\algorithmicrequire}{\textbf{BEGIN}}
\renewcommand{\algorithmicensure}{\textbf{END}}
\REQUIRE \hspace{0.1cm} \\   
    \textbf{Process}: \\
    \FOR{$ro = 0$ to $(D_{ro}-1)$}
    \FOR{$su = 0$ to $(D_{su}-1)$}
    \FOR{$ba = 0$ to $(D_{ba}-1)$}
    \FOR{$co = 0$ to $(D_{co}-1)$}
      \IF{$Drate\_col \leq Drate\_col_{max}$}
          \STATE $DRAM[ba, su, ro, co] \leftarrow  FAM(weight\_b)$;
      \ENDIF
    \ENDFOR
    \ENDFOR
    \ENDFOR
    \ENDFOR
    \RETURN $DRAM$;
\ENSURE 
\end{algorithmic}
\end{algorithm}
\setlength{\textfloatsep}{2pt}
\begin{algorithm}[hbtp]
\footnotesize
\caption{\color{black} The proposed mapping for SRAM buffer}
\label{Alg_SRAMmapping}
\begin{algorithmic}[1]
\renewcommand{\algorithmicrequire}{\textbf{INPUT:}}
\renewcommand{\algorithmicensure}{\textbf{OUTPUT:}}
\REQUIRE \textbf{(1)} SRAM ($SRAM$), number of bank ($S_{ba}$), number of row-per- bank ($S_{ro}$); // the number of column-per-row = the bitwidth of a word \\
\textbf{(2)} Fault rate of an SRAM row ($Srate\_row$), maximum tolerable fault rate for an SRAM row ($Srate\_row_{max}$);\\
\textbf{(3)} Weight bits ($weight\_b$); \\
\textbf{(4)} Fault-aware mapping ($FAM$); // either FAM1 or FAM2\\
\ENSURE SRAM ($SRAM$); \\
\vspace{0.1cm}
\renewcommand{\algorithmicrequire}{\textbf{BEGIN}}
\renewcommand{\algorithmicensure}{\textbf{END}}
\REQUIRE \hspace{0.1cm} \\   
    \textbf{Process}: \\
    \FOR{$ro = 0$ to $(S_{ro}-1)$}
    \FOR{$ba = 0$ to $(S_{ba}-1)$}
      \IF{$Srate\_row \leq Srate\_row_{max}$}
          \STATE $SRAM[ba, ro] \leftarrow  FAM(weight\_b)$;
      \ENDIF
    \ENDFOR
    \ENDFOR
    \RETURN $SRAM$;
\ENSURE 
\end{algorithmic}
\end{algorithm}
\setlength{\textfloatsep}{4pt}

\textbf{Note:} The proposed FAM techniques (FAM1 and FAM2) do not require retraining, thereby making them suitable for energy-efficient and fault-tolerant SNN processing, especially in the case where the training dataset is not fully available. 
Consequently, these techniques can also improve the yield and reduce the per-unit-cost of SNN accelerators (hardware chips). 

\subsection{Fault-aware Training-and-Mapping (FATM)}
\label{Sec_ReSpawn_FMT}
\vspace{-0.1cm}

If the training dataset is fully available, users can decide if they want to perform fault-mitigation techniques without training, like our FAM techniques (Section~\ref{Sec_ReSpawn_FM}), or fault-aware training (FAT)\footnote{Fault-aware training (FAT) is a widely used technique for improving the fault-tolerance of neural networks, by incorporating the information of faults in the training process \cite{Ref_Koppula_EDEN_MICRO19}\cite{Ref_Putra_SparkXD_arXiv21}\cite{Ref_Zhang_PermanentFaults_VTS18}.}.
Our experimental results in Fig.~\ref{Fig_ObserveArchAccDrop}(c) show that, the FAT technique can improve the SNN fault tolerance.
Toward this, \textit{ReSpawn framework also provides FAT-based solutions to improve the SNN fault tolerance on top of the proposed FAM techniques; so-called fault-aware training-and-mapping (FATM)}. 
For conciseness, the FAT with FAM1 mapping is referred to as the FATM1, and the FAT with FAM2 mapping is referred to as the FATM2.
The proposed FATM is performed through the following mechanisms. 
\begin{enumerate}[leftmargin=*]
    \item We employ the FAM technique (FAM1 or FAM2) on the given SNN model, to minimize the negative impacts of faults on the weights. 
    This results in the SNN model whose weights have been minimally affected by the faults (i.e., the FAM-improved SNN model).
    \item Afterward, we perform training to the FAM-improved SNN model 
    through the following steps.
    \begin{itemize}
        \item \textbf{Step-1:} The faults are generated for different rates, based on the SNN fault tolerance analysis in Section~\ref{Sec_ReSpawn_SNNftAnalysis}. 
        \item \textbf{Step-2:} The generated faults are injected into locations in DRAM and weight buffer, thereby causing the weight bits stored in these locations to flip. 
        \item \textbf{Step-3:} We train the SNN model while considering fault rates that do not cause accuracy drop (fault rates from region-\circled{A} in Fig.~\ref{Fig_FaultToleranceAnalysis}) and are close to region-\circled{B}.  
        It makes the model adapting to high fault rates safely, without causing accuracy to decrease, and with less training time, since smaller fault rates are not considered. 
        %
        \item \textbf{Step-4:} Afterward, we carefully train the SNN model while considering fault rates that cause notable accuracy drop, i.e., fault rates from region-\circled{B}, by incrementally increasing the fault rates of DRAM and weight buffer, after each training epoch.
        \item \textbf{Step-5:} Training is terminated when the network faces accuracy saturation or degradation. 
        The final SNN model is selected from the trained model that is saved in the previous training epoch.
    \end{itemize}
\end{enumerate}

\add{The proposed FAM (FAM1 and FAM2) and FATM (FATM1 and FATM2) techniques are applicable for different memory technologies (like CMOS, RRAM, etc.) since they consider bit-level fault mitigation, which is suitable for bit-level data storage in each memory cell. 
Therefore, the possible extension to the ReSpawn framework is by considering the multi-level cell characteristics into its optimization process.}

\section{Evaluation Methodology}
\label{Sec_EvalMethod}

\begin{figure}[hbtp]
\centering
\includegraphics[width=\linewidth]{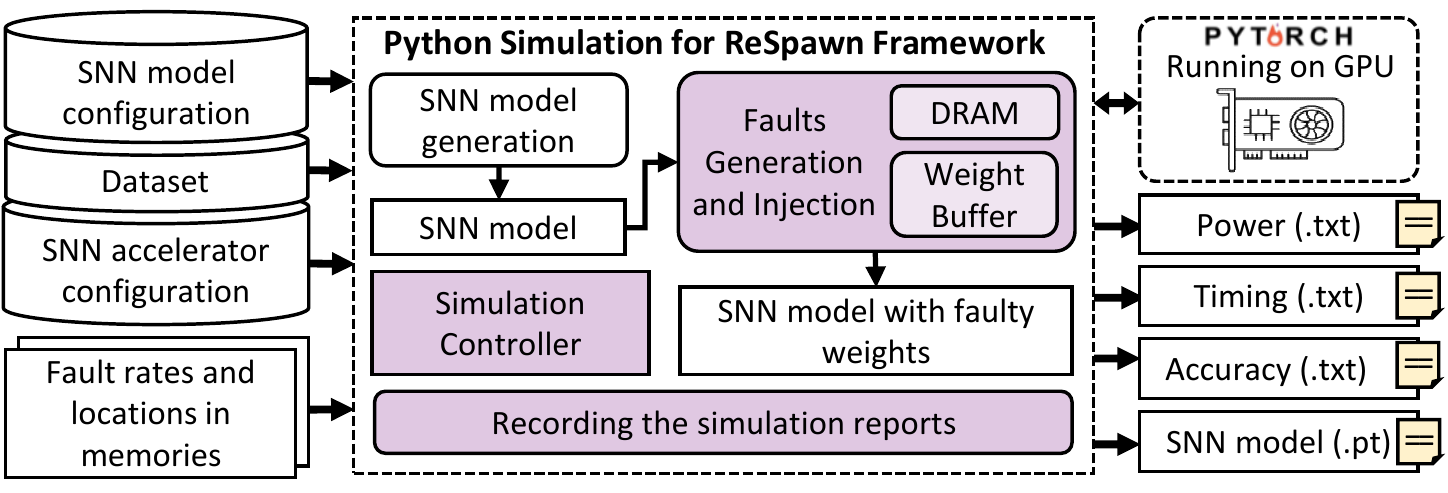}
\vspace{-0.7cm}
\caption{Experimental setup and tool flow.}
\label{Fig_ExpSetup}
\end{figure}

The experimental setup for evaluating ReSpawn framework is illustrated in Fig.~\ref{Fig_ExpSetup}. 
Following is the detailed information of the evaluation methodology, comprising the scenarios for experiments and comparisons.

\add{For the network architecture, we use the single-layer fully- connected SNN, like the network in Fig.~\ref{Fig_ObserveTopoMemAcc}(a), with a different number of neurons (i.e., 100, 400, 900, 1600, 2500, and 3600) to show the generality of the ReSpawn}, which we refer them to as Net100, Net400, Net900, Net1600, Net2500, and Net3600, respectively.
\add{We consider this network as it provides robustness when performing different variants learning rules \cite{Ref_Diehl_STDPmnist_FNCOM15}, thereby it is representative for evaluation}.
Meanwhile, for the comparison partners, we consider two designs: (1) the baseline SNN model without any fault-mitigation technique, and (2) the SNN model with the FAT technique.
We compare these designs against our ReSpawn techniques (i.e., FAM1, FAM2, FATM1, and FATM2) on the MNIST dataset.

\textbf{Fault Injection:} 
We generate memory faults based on the fault modeling described in Section~\ref{Sec_ErrorModelMemories}.  
Afterward, we inject these faults into the locations in DRAM and weight buffer to represent the faulty memory cells, and the data bits in these cells are flipped. 
For ReSpawn, we employ mapping policies from Alg.~\ref{Alg_DRAMmapping} and Alg.~\ref{Alg_SRAMmapping}, while for the baseline, we store the weights in subsequent addresses in a DRAM bank.

\textbf{Accuracy Evaluation:}
To evaluate the accuracy of SNNs, we use Python-based simulations \cite{Ref_Hazan_BindsNET_FNINF18} that run on GPGPU, i.e., Nvidia RTX 2080 Ti, while considering an SNN accelerator architecture that follows the design of \cite{Ref_Frenkel_ODIN_TBCAS19} with 8-bit precision of weights, a DDR3-1600 2Gb DRAM, and a 32KB weight buffer, as shown in Fig.~\ref{Fig_ObserveArchAccDrop}(a). 
We use 8-bit precision as it has a sufficient range of values to represent the SNN weights \cite{Ref_Putra_FSpiNN_TCAD20}.

\textbf{Energy Evaluation:} 
We consider the approach of \cite{Ref_Han_DeepCompress_ICLR16} for estimating the SNN processing energy of an SNN model, i.e., by leveraging the information of processing power that is obtained through \textit{nvidia-smi} utility, and its processing time. 
We perform the energy evaluation for different scenarios, i.e., the fault-mitigation techniques without retraining (i.e., FAM1 and FAM2) and with retraining (i.e., FAT, FATM1, and FATM2).


\section{Results and Discussions}
\label{Sec_Results}

\subsection{Maintaining the Accuracy}
\label{Sec_Results_Accuracy}

\begin{figure*}[t]
\centering
\includegraphics[width=\linewidth]{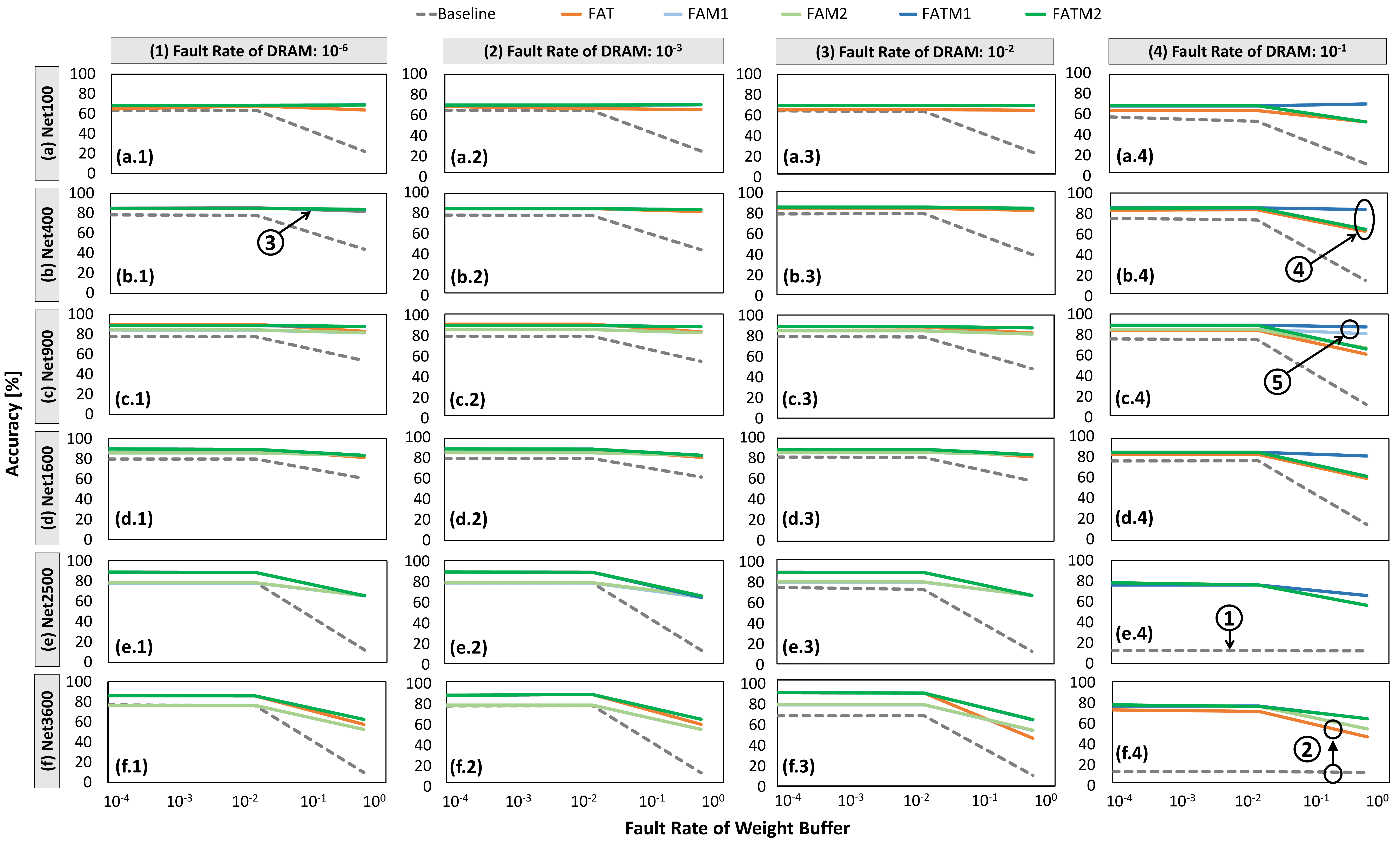}
\vspace{-0.9cm}
\caption{Accuracy achieved by different techniques, i.e., the baseline, the fault-aware training (FAT), our fault-aware mapping (FAM1 and FAM2), and our fault-aware training-and-mapping (FATM1 and FATM2), across different sizes of network, i.e., (a) Net100, (b) Net400, (c) Net900, (d) Net1600, (e) Net2500, and (f) Net3600, as well as different fault rates in DRAM and weight buffer.}
\label{Fig_Results_Accuracy}
\vspace{-0.5cm}
\end{figure*}

Fig.~\ref{Fig_Results_Accuracy} presents the experimental results on the accuracy of different fault-mitigation techniques, i.e., the baseline, the FAT, our FAM techniques (FAM1 and FAM2), and our FATM techniques (FATM1 and FATM2). 

We observe that the baseline is susceptible to accuracy degradation when the SNN is run under the presence of faults, as these faults alter the weights and affect the output of the SNN model. 
The accuracy degradation is more evident in the scenarios where high fault rates are observed, as shown by label-\circled{1}.
The FAT technique improves the SNN fault tolerance compared to the baseline across all evaluation scenarios, as the FAT-improved SNN model has a better capability for adapting to the presence of faults, as shown by label-\circled{2}. 
However, the FAT technique may offer limited accuracy improvements, since it does not substantially eliminate the negative impact of faults on the significant bits of weights, and its performance depends on the effectiveness of the training strategy.
On the other hand, our FAM techniques (FAM1 and FAM2) can achieve comparable accuracy compared to the FAT without retraining, as shown by label-\circled{3}. 
They improve the accuracy by up to 61\%, 70\%, 70\%, 67\%, 53\%, and 43\% for Net100, Net400, Net900, Net1600, Net2500, and Net3600 respectively, compared to the baseline. 
The reason is that, the main idea of our FAM1 and FAM2 is to eliminate the impact of faults on the significant bits of weights through simple bit-shuffling, thereby maintaining the value of weights as close as possible to the weights that are trained in an ideal condition, i.e., environment without faults. 
These results show that \textit{our FAM techniques (FAM1 and FAM2) are effective for fault-mitigation techniques in SNNs, especially in the case where the training dataset is not fully available}. 
Moreover, these techniques can enhance the yield, thereby decreasing the per-unit-cost of SNN chips.

We also observe that the FAM1 has better performance than the FAM2, as it consistently obtains high accuracy across all evaluation scenarios, while the performance of the FAM2 may still be affected by cases that have high fault rates, as shown by label-\circled{4}. 
The reason is that, the FAM1 minimizes the impact of faults on the significant bits of weights from each memory. 
Meanwhile, the FAM2 minimizes the impact of faults on the significant bits of weights considering the integrated fault map from multiple memories, which may be sub-optimal for each memory. 
Furthermore, we observe that our FATM techniques can further improve the SNN fault tolerance from the FAM techniques (shown in label-\circled{5}), since the training is performed to the model whose weights are already minimally affected by the faults. 
The FATM techniques improve the accuracy by up to 61\%, 70\%, 76\%, 67\%, 53\%, and 53\% for Net100, Net400, Net900, Net1600, Net2500, and Net3600 respectively, compared to the baseline. 
These results show that, \textit{our FATM techniques (FATM1 and FATM2) can further improve the SNN fault tolerance if the training dataset is fully available}. 


\begin{figure*}[t]
\centering
\includegraphics[width=0.92\linewidth]{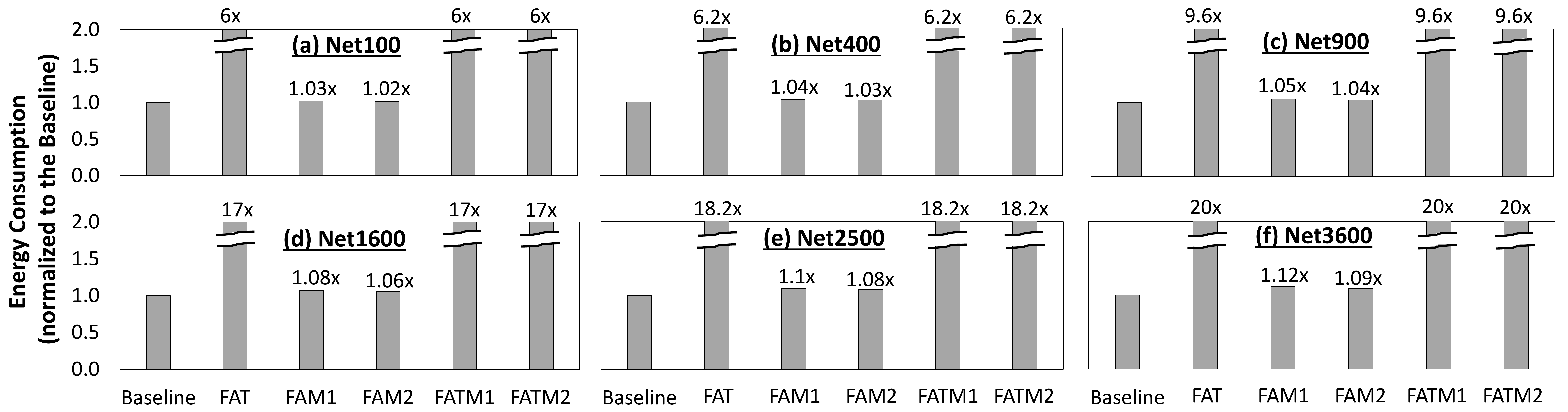}
\vspace{-0.3cm}
\caption{Normalized energy consumption of different fault-mitigation techniques (i.e., the baseline, the FAT, our FAM1 and FAM2, as well as our FATM1 and FATM2) across different sizes of network: (a) Net100, (b) Net400, (c) Net900, (d) Net1600, (e) Net2500, and (f) Net3600.} 
\label{Fig_Results_Energy}
\vspace{-0.4cm}
\end{figure*}

\subsection{Reducing the Energy Consumption}
\label{Sec_Results_Energy}
\vspace{-0.1cm}

Fig.~\ref{Fig_Results_Energy} presents the experimental results on the energy consumption of different fault-mitigation techniques, i.e., the baseline, the FAT, our FAM1, FAM2, FATM1 and FATM2. 
For the training-based solutions (FAT, FATM1, and FATM2), the energy consumption is evaluated when performing one training epoch over 60k samples from the full MNIST training set.
For the solutions without training (FAM1 and FAM2), the energy consumption is evaluated when performing a test over the 10k samples from the full MNIST test set. 
\textit{This evaluation scenario aims at showing how much energy the training-based solutions incur, compared to the solutions without training.}

We observe that, the FAT technique consumes high energy across different network sizes, since it requires the training process. 
Moreover, a larger-sized network incurs higher power and processing time, and thereby higher energy.
If we consider one training epoch (i.e., running 60k samples from the full MNIST training set), the FAT incurs about 6x-20x energy for Net100-Net3600, compared to the baseline. 
This condition can be exacerbated by the fact that the energy consumption is increased if the FAT requires multiple training epochs. 
Here, the FATM1 and FATM2 techniques face the same issues due to the training-based approach.
On the other hand, our FAM1 technique only incurs 1.03x-1.12x energy, and our FAM2 technique only incurs 1.02x-1.09x energy for Net100-Net3600 compared to the baseline, when running 10k samples from a full MNIST test set. 
Moreover, the energy efficiency of the FAM1 and the FAM2 can be better if the number of samples in the inference phase is higher.    
The reason is that, the mapping patterns in the FAM1 and the FAM2 need to be generated only once, before running the inference, therefore the energy overhead is negligible considering the huge number of samples to be processed in the inference phase. 
We also observe that the FAM2 incurs slightly less energy compared to the FAM1, as the FAM2 only considers one integrated fault map for its mapping operations while the FAM1 considers multiple fault maps for its mapping operations, thereby incurring fewer operations and processing energy.  
These results show that, \textit{our FAM techniques (FAM1 and FAM2) have high potential as the energy-efficient fault-mitigation techniques for SNNs}, as they maintain high accuracy with minimum energy overhead.

\vspace{-0.1cm}
\section{Conclusion}
\label{Sec_Conclusion}
\vspace{-0.1cm}

We propose ReSpawn framework for mitigating the faults in the off-chip and on-chip weight memories for SNN-based systems. 
Our ReSpawn employs SNN fault tolerance analysis, fault-aware mapping, and fault-aware mapping-and-training. 
The experimental results show that, ReSpawn with fault-aware mapping improves the accuracy by up to 70\% for a 900-neuron network without retraining. 
Therefore, our work enhances the SNN fault tolerance with minimum energy overhead, thereby potentially improving the yield of SNN hardware chips. 


\vspace{-0.1cm}
\section{Acknowledgment}
\vspace{-0.1cm}

This work was partly supported by Intel Corporation through Gift funding for the project ”Cost-Effective Dependability for Deep Neural Networks and Spiking Neural Networks”, and by Indonesia Endowment Fund for Education (LPDP) through Graduate Scholarship Program. 

\bibliographystyle{IEEEtran}
\bibliography{bibliography}

\begin{thebibliography}{10}
\providecommand{\url}[1]{#1}
\csname url@samestyle\endcsname
\providecommand{\newblock}{\relax}
\providecommand{\bibinfo}[2]{#2}
\providecommand{\BIBentrySTDinterwordspacing}{\spaceskip=0pt\relax}
\providecommand{\BIBentryALTinterwordstretchfactor}{4}
\providecommand{\BIBentryALTinterwordspacing}{\spaceskip=\fontdimen2\font plus
\BIBentryALTinterwordstretchfactor\fontdimen3\font minus
  \fontdimen4\font\relax}
\providecommand{\BIBforeignlanguage}[2]{{%
\expandafter\ifx\csname l@#1\endcsname\relax
\typeout{** WARNING: IEEEtran.bst: No hyphenation pattern has been}%
\typeout{** loaded for the language `#1'. Using the pattern for}%
\typeout{** the default language instead.}%
\else
\language=\csname l@#1\endcsname
\fi
#2}}
\providecommand{\BIBdecl}{\relax}
\BIBdecl

\bibitem{Ref_Pfeiffer_DLSNN_FNINS18}
M.~Pfeiffer and T.~Pfeil, ``Deep learning with spiking neurons: Opportunities
  and challenges,'' \emph{Frontiers in Neuroscience}, vol.~12, 2018.

\bibitem{Ref_Tavanaei_DLSNN_Neunet18}
A.~Tavanaei \emph{et~al.}, ``Deep learning in spiking neural networks,''
  \emph{Neural Networks}, vol. 111, pp. 47--63, 2019.

\bibitem{Ref_Putra_FSpiNN_TCAD20}
R.~V.~W. {Putra} and M.~{Shafique}, ``Fspinn: An optimization framework for
  memory-efficient and energy-efficient spiking neural networks,'' \emph{IEEE
  TCAD}, vol.~39, no.~11, pp. 3601--3613, 2020.

\bibitem{Ref_Putra_SpikeDyn_arXiv21}
R.~V.~W. Putra and M.~Shafique, ``Spikedyn: A framework for energy-efficient
  spiking neural networks with continual and unsupervised learning capabilities
  in dynamic environments,'' \emph{arXiv preprint arXiv:2103.00424}, 2021.

\bibitem{Ref_Akopyan_TrueNorth_TCAD15}
F.~{Akopyan} \emph{et~al.}, ``Truenorth: Design and tool flow of a 65 mw 1
  million neuron programmable neurosynaptic chip,'' \emph{IEEE TCAD}, vol.~34,
  no.~10, pp. 1537--1557, 2015.

\bibitem{Ref_Roy_PEASE_ISLPED17}
A.~{Roy} \emph{et~al.}, ``A programmable event-driven architecture for
  evaluating spiking neural networks,'' in \emph{Proc. of ISLPED}, July 2017,
  pp. 1--6.

\bibitem{Ref_Sen_ApproxSNN_DATE17}
S.~{Sen} \emph{et~al.}, ``Approximate computing for spiking neural networks,''
  in \emph{Proc. of DATE}, March 2017, pp. 193--198.

\bibitem{Ref_Davies_Loihi_MM18}
M.~{Davies} \emph{et~al.}, ``Loihi: A neuromorphic manycore processor with
  on-chip learning,'' \emph{IEEE Micro}, vol.~38, no.~1, pp. 82--99, Jan 2018.

\bibitem{Ref_Frenkel_ODIN_TBCAS19}
C.~{Frenkel} \emph{et~al.}, ``A 0.086-mm$^2$ 12.7-pj/sop 64k-synapse 256-neuron
  online-learning digital spiking neuromorphic processor in 28-nm cmos,''
  \emph{IEEE TBCAS}, vol.~13, no.~1, pp. 145--158, Feb 2019.

\bibitem{Ref_Frenkel_MorphIC_TBCAS19}
C.~Frenkel \emph{et~al.}, ``Morphic: A 65-nm 738k-synapse/mm$^2$ quad-core
  binary-weight digital neuromorphic processor with stochastic spike-driven
  online learning,'' \emph{IEEE TBCAS}, vol.~13, no.~5, 2019.

\bibitem{Ref_Koren_Yield_IEEE98}
I.~Koren and Z.~Koren, ``Defect tolerance in vlsi circuits: techniques and
  yield analysis,'' \emph{Proc. of the IEEE}, vol.~86, no.~9, pp. 1819--1838,
  1998.

\bibitem{Ref_Chang_Voltron_POMACS17}
K.~K. Chang \emph{et~al.}, ``Understanding reduced-voltage operation in modern
  dram devices: Experimental characterization, analysis, and mechanisms,''
  \emph{ACM POMACS}, vol.~1, no.~1, Jun. 2017.

\bibitem{Ref_Ganapathy_UnreliableMem_DAC15}
S.~Ganapathy \emph{et~al.}, ``Mitigating the impact of faults in unreliable
  memories for error-resilient applications,'' in \emph{Proc. of DAC}, 2015.

\bibitem{Ref_Vadlamani_DMR_DATE20}
R.~{Vadlamani} \emph{et~al.}, ``Multicore soft error rate stabilization using
  adaptive dual modular redundancy,'' in \emph{Proc. of DATE}, 2010, pp.
  27--32.

\bibitem{Ref_Lyons_TMR_IBM62}
R.~E. Lyons and W.~Vanderkulk, ``The use of triple-modular redundancy to
  improve computer reliability,'' \emph{IBM J. Res. Dev.}, vol.~6, no.~2, 1962.

\bibitem{Ref_Sze_ECCs_USPatent00}
H.~Y. Sze, ``Circuit and method for rapid checking of error correction codes
  using cyclic redundancy check,'' Jul.~18 2000, uS Patent 6,092,231.

\bibitem{Ref_Vatajelu_ReliabilitySNN_VTS19}
E.-I. Vatajelu \emph{et~al.}, ``Special session: Reliability of hardware-
  implemented spiking neural networks (snn),'' in \emph{Proc. of VTS}, 2019.

\bibitem{Ref_Sayed_NeuronFaultModel_IOLTS20}
S.~A. El-Sayed \emph{et~al.}, ``Spiking neuron hardware-level fault modeling,''
  in \emph{Proc. of IOLTS}, 2020, pp. 1--4.

\bibitem{Ref_Spyrou_NeuronFT_DATE21}
T.~Spyrou \emph{et~al.}, ``Neuron fault tolerance in spiking neural networks,''
  in \emph{Proc. of DATE}, 2021.

\bibitem{Ref_Venceslai_NeuroAttack}
V.~Venceslai \emph{et~al.}, ``Neuroattack: Undermining spiking neural networks
  security through externally triggered bit-flips,'' in \emph{Proc. of IJCNN},
  2020, pp. 1--8.

\bibitem{Ref_Schuman_ResilinceSNN_IJCNN20}
C.~D. Schuman \emph{et~al.}, ``Resilience and robustness of spiking neural
  networks for neuromorphic systems,'' in \emph{Proc. of IJCNN}, 2020, pp.
  1--10.

\bibitem{Ref_Rastogi_AstrocytesSTDP_FNINS21}
M.~Rastogi \emph{et~al.}, ``On the self-repair role of astrocytes in stdp
  enabled unsupervised snns,'' \emph{Frontiers in Neuroscience}, vol.~14, p.
  1351, 2021.

\bibitem{Ref_Koppula_EDEN_MICRO19}
S.~Koppula \emph{et~al.}, ``Eden: Enabling energy-efficient, high-performance
  deep neural network inference using approximate dram,'' in \emph{Proc. of
  MICRO}, 2019, p. 166–181.

\bibitem{Ref_Putra_SparkXD_arXiv21}
R.~V.~W. Putra \emph{et~al.}, ``Sparkxd: A framework for resilient and
  energy-efficient spiking neural network inference using approximate dram,''
  \emph{arXiv preprint arXiv:2103.00421}, 2021.

\bibitem{Ref_Gautrais_SpikeCoding_Bio98}
J.~Gautrais and S.~Thorpe, ``Rate coding versus temporal order coding: a
  theoretical approach,'' \emph{Biosystems}, vol.~48, no.~1, pp. 57--65, 1998.

\bibitem{Ref_Thorpe_RankOrder_Springer98}
S.~Thorpe and J.~Gautrais, ``Rank order coding,'' in \emph{Computational
  neuroscience}.\hskip 1em plus 0.5em minus 0.4em\relax Springer, 1998, pp.
  113--118.

\bibitem{Ref_Park_BurstSNN_DAC19}
S.~Park \emph{et~al.}, ``Fast and efficient information transmission with burst
  spikes in deep spiking neural networks,'' in \emph{Proc. of DAC}, 2019,
  p.~53.

\bibitem{Ref_Park_T2FSNN_DAC20}
------, ``T2fsnn: Deep spiking neural networks with time-to-first-spike
  coding,'' in \emph{Proc. of DAC}, 2020.

\bibitem{Ref_Putra_QSpiNN_arXiv21}
R.~V.~W. Putra and M.~Shafique, ``Q-spinn: A framework for quantizing spiking
  neural networks,'' \emph{arXiv preprint arXiv:2107.01807}, 2021.

\bibitem{Ref_Cesar_ReviewFTinNN_Access17}
C.~Torres-Huitzil and B.~Girau, ``Fault and error tolerance in neural networks:
  A review,'' \emph{IEEE Access}, vol.~5, pp. 17\,322--17\,341, 2017.

\bibitem{Ref_Hanif_RobustML_IOLTS18}
M.~A. {Hanif} \emph{et~al.}, ``Robust machine learning systems: Reliability and
  security for deep neural networks,'' in \emph{Proc. of IOLTS}, 2018, pp.
  257--260.

\bibitem{Ref_Zhang_RobustML_DAC19}
J.~J. Zhang \emph{et~al.}, ``Building robust machine learning systems: Current
  progress, research challenges, and opportunities,'' in \emph{Proc. of DAC},
  2019, pp. 1--4.

\bibitem{Ref_Shafique_RobustML_DnT20}
M.~Shafique \emph{et~al.}, ``Robust machine learning systems: Challenges,
  current trends, perspectives, and the road ahead,'' \emph{IEEE Design \&
  Test}, vol.~37, no.~2, pp. 30--57, 2020.

\bibitem{Ref_Zhang_PermanentFaults_VTS18}
J.~J. Zhang \emph{et~al.}, ``Analyzing and mitigating the impact of permanent
  faults on a systolic array based neural network accelerator,'' in \emph{Proc.
  of VTS}, 2018, pp. 1--6.

\bibitem{Ref_Abdullah_SalvageDNN_RSTA20}
M.~A. Hanif and M.~Shafique, ``Salvagednn: salvaging deep neural network
  accelerators with permanent faults through saliency-driven fault-aware
  mapping,'' \emph{Philosophical Trans. of the Royal Society A}, vol. 378, no.
  2164, p. 20190164, 2020.

\bibitem{Ref_Hassan_SoftMC_HPCA17}
H.~Hassan \emph{et~al.}, ``Softmc: A flexible and practical open-source
  infrastructure for enabling experimental dram studies,'' in \emph{Proc. of
  HPCA}, 2017, pp. 241--252.

\bibitem{Ref_Krithivasan_SpikeBundle_ISLPED19}
S.~{Krithivasan} \emph{et~al.}, ``Dynamic spike bundling for energy-efficient
  spiking neural networks,'' in \emph{Proc. of ISLPED}, July 2019, pp. 1--6.

\bibitem{Ref_Ghose_DRAMworkload_Sigmetrics19}
S.~Ghose \emph{et~al.}, ``Demystifying complex workload-dram interactions: An
  experimental study,'' in \emph{Proc. of SIGMETRICS}, 2019, pp. 93--93.

\bibitem{Ref_Putra_ROMANet_TVLSI21}
R.~V.~W. Putra \emph{et~al.}, ``Romanet: Fine-grained reuse-driven off-chip
  memory access management and data organization for deep neural network
  accelerators,'' \emph{IEEE TVLSI}, vol.~29, no.~4, pp. 702--715, 2021.

\bibitem{Ref_Kim_SALP_ISCA12}
Y.~{Kim} \emph{et~al.}, ``A case for exploiting subarray-level parallelism
  (salp) in dram,'' in \emph{Proc. of ISCA}, June 2012.

\bibitem{Ref_Putra_DRMap_DAC20}
R.~V.~W. {Putra} \emph{et~al.}, ``Drmap: A generic dram data mapping policy for
  energy-efficient processing of convolutional neural networks,'' in
  \emph{Proc. of DAC}, 2020, pp. 1--6.

\bibitem{Ref_Diehl_STDPmnist_FNCOM15}
P.~Diehl and M.~Cook, ``Unsupervised learning of digit recognition using
  spike-timing-dependent plasticity,'' \emph{Frontiers in Computational
  Neuroscience}, vol.~9, p.~99, 2015.

\bibitem{Ref_Hazan_BindsNET_FNINF18}
H.~Hazan \emph{et~al.}, ``Bindsnet: A machine learning-oriented spiking neural
  networks library in python,'' \emph{Frontiers in Neuroinformatics}, 2018.

\bibitem{Ref_Han_DeepCompress_ICLR16}
S.~Han \emph{et~al.}, ``Deep compression: Compressing deep neural networks with
  pruning, trained quantization and huffman coding,'' \emph{arXiv preprint
  arXiv:1510.00149}, 2015.

\end{thebibliography}

\end{document}